\documentclass[conference]{IEEEtran}
\IEEEoverridecommandlockouts
\usepackage{cite}
\usepackage{hyperref}
\usepackage[T1]{fontenc}
\hypersetup{colorlinks=true,linkcolor=black,citecolor=blue,filecolor=black,urlcolor=blue}
\usepackage[euler]{textgreek}
\usepackage{amsmath,amssymb,amsfonts}
\usepackage{algorithmic}
\usepackage{graphicx}
\usepackage{textcomp}
\usepackage{xcolor}
\usepackage{tabularx}
\usepackage{xspace}
\usepackage{subcaption}
\usepackage{svg}
\usepackage{amsmath}
\usepackage[normalem]{ulem}

\def\BibTeX{{\rm B\kern-.05em{\sc i\kern-.025em b}\kern-.08em
    T\kern-.1667em\lower.7ex\hbox{E}\kern-.125emX}}
\begin{document}

\title{A Recommender System for Scientific Datasets and Analysis Pipelines}
\newcommand{\TG}[1]{\noindent{\color{blue}[\textsc{From Tristan:} #1]}\xspace}
\newcommand{\GK}[2]{\noindent{\color{brown}\sout{#1}\xspace\color{violet}\textit{#2}}\xspace}

\newcommand{\MM}[1]{\noindent{\color{blue}[\textsc{From Mandana:} #1]}\xspace}

\newcommand{\WIP}[1]{\noindent{\color{gray} [WIP: #1]}\xspace}


\author{
    \IEEEauthorblockN{Mandana Mazaheri$^1$, Gregory Kiar$^2$, Tristan Glatard$^1$}
    \IEEEauthorblockA{$^1$Department of Computer Science and Software Engineering, Concordia University, Montreal, Canada\\
    $^2$Center for the Developing Brain, Child Mind Institute, New York, NY, USA}
}

\maketitle
\begin{abstract}
Scientific datasets and analysis pipelines are increasingly being shared
publicly in the interest of open science.
However, mechanisms are lacking to reliably identify which pipelines
and datasets can appropriately be used together. Given the increasing number of high-quality public datasets and
pipelines, this lack of clear compatibility threatens the
findability and reusability of these resources. We investigate
the feasibility of a collaborative filtering system to recommend pipelines
and datasets based on provenance records from previous executions.
We evaluate our system using datasets and pipelines extracted from the
Canadian Open Neuroscience Platform, a national initiative for open
neuroscience. The recommendations provided by our system (AUC$=0.83$) are
significantly better than chance and outperform recommendations made by
domain experts using their previous knowledge as well as pipeline and dataset descriptions (AUC$=0.63$). In particular, domain experts often neglect
low-level technical aspects of a pipeline-dataset interaction, such as the level of pre-processing, which are
captured by a provenance-based system. We conclude that provenance-based
pipeline and dataset recommenders are feasible and beneficial to
the sharing and usage of open-science resources. Future 
work will focus on the collection of more
comprehensive provenance traces, and on deploying the system in production.

\end{abstract}

\begin{IEEEkeywords}
Scientific dataset, recommendation, collaborative filtering
\end{IEEEkeywords}

\section{Introduction}

Open science has emerged as a framework to improve the quality of
scientific analyses, ideally leading to Findable, Accessible, Interoperable, and
Reusable (FAIR~\cite{wilkinson2016fair}) datasets and analysis pipelines. In
neuroscience, our main application domain of interest, platforms have
emerged to facilitate the sharing of datasets and pipelines, including
OpenNeuro~\cite{gorgolewski2017openneuro}, NeuroImaging Tools and Resources Collaboratory (NITRC~\cite{kennedy2016nitrc}), and the Canadian Open
Neuroscience Platform~\cite{conp}. However, while public
datasets and pipelines proliferate, researchers remain unassisted in
creating relevant analyses from these resources that therefore remain
largely underutilized. Identifying the set of analysis pipelines that may
be relevantly applied to a given dataset, or conversely the list of datasets that may be processed by
a given pipeline, remains challenging. In this paper we investigate
the development and use of recommender systems to address this issue.


The past decades witnessed the adoption of recommender systems as the major
 technology to help customers navigate the abundant product offerings of
 online retail platforms. Two main
 recommending strategies emerged: content-based strategies, which match content-rich item profiles with user profiles~\cite{pazzani2007content},
 and collaborative filtering, which recommends items to a given user based on those
 selected by users with similar preferences~\cite{rajaraman2011mining}.
 Both strategies have been successfully applied and have their own strengths and weaknesses.
 
 While these techniques have largely been employed in the space of retail
 content delivery, we evaluate the feasibility of matching scientific pipelines and datasets
 using existing recommender system techniques. We focus on collaborative
 filtering approaches, since content-based methods would require extensive
 annotations about pipelines and datasets which are not broadly available
 despite on-going
 efforts~\cite{NeuroimagingDataModel,sansone2017dats,DATSDocumentation}.
 However, collaborative filtering requires an affinity measure
 between users and items. To define such a measure between pipelines and
 datasets, we rely on past execution outcomes (e.g. exit status) available through provenance records.

The compatibility of a pipeline with a dataset depends on
semantic, syntactic, and infrastructural factors. Semantic factors refer to
the content of datasets and analyses. For instance, a pipeline developed
for the segmentation of brain Magnetic Resonance Images (MRIs) would not produce
meaningful results on other image types. Syntactic factors refer to file
formats and dataset organization. For instance, brain images are commonly
stored using the NIfTI~\cite{larobina2014medical} or
MINC~\cite{vincent2016minc} formats and pipelines developed to ingest one
format may not apply to others. In addition, the multiple files and
directories composing a dataset are increasingly structured using the Brain
Imaging Data Structure (BIDS~\cite{bids}) which is required by some
pipelines while other ones use their own structure or neglect structure entirely and require explicit pointers to specific files. Finally,
infrastructural factors refer to the availability of pipelines
and datasets which can be functionally deployed. For instance, some pipelines may require the loading of an
entire dataset in memory, which may not be feasible for large datasets. All
these levels must be considered in pipeline or dataset recommendations.

Provenance is a key concept in computational analyses, referring to the
detailed description of data transformations. Provenance records typically include
information about the input data, the analysis
software and parameters, the execution context and finally, the execution outcomes. In neuroimaging, the NeuroImaging Data Model (NIDM~\cite{maumet2016sharing}) project proposed domain-specific
formats and tools based on standards from the W3C PROV working
group~\cite{missier2013w3c}. In its current form, our
recommender system merely relies on execution outcomes (summarized via ``exit codes")
extracted from these or similar provenance records.

To summarize, this paper makes the following contributions:
\begin{itemize}
\item  Describes and presents a provenance-based recommender system for scientific
pipelines and datasets;
\item Evaluates the system for datasets and pipelines from the Canadian
Open Neuroscience Platform;
\item Compares the system against domain expert recommendations.
\end{itemize}

\section{Related Work}
    
Systems have been used to recommend software in various contexts such as workflow composition and algorithm selection. However, as explained below, our context is slightly different since we aim at recommending analyses that are applicable to a given dataset. In neuroimaging, existing platforms focus on finding and reusing pipelines and datasets but do not include any recommender engine.
    
\subsection{Workflow composition}

Recommender systems have been described to assist users with workflow composition, in particular to identify candidate software components for a given workflow. For example, the Galaxy tool recommender~\cite{kumar2021tool} recommends possible workflow components using a deep learning model trained on $18,000$ bioinformatics workflows from the European Galaxy server. Recommendations depend on the definition and organization of all the tools in the workflow (so-called ``higher-order workflow dependencies") instead of focusing only on the most recently added workflow components. Workflow dependencies are learned using Recurrent Neural Networks, resulting in a mean accuracy of 98\%. The system is available for Galaxy users as an extension. 

Previous approaches to assisted workflow composition included loose programming, initially proposed in the PROPHETS system~\cite{lamprecht2010synthesis,naujokat2012loose,lamprecht2013user}. Loose programming enables workflow developers to program using concepts rather than accurate procedural code. Loose programming exposes to the workflow developers semantic annotations describing the functionalities of workflow components. Workflow developers can then assemble such components without having to care about correct typing, interface compatibility, platform parameters or other technical details that are taken care of through subsequent validations. PROPHETS was applied to various bioinformatics use cases, including mass spectrometry-based proteomics~\cite{palmblad2019automated}.


The WINGS (Workflow INstance Generation and Specialization) system~\cite{gil2010wings} uses AI planning and semantic reasoners to assist users in creating workflows while validating that the workflows comply with the requirements of the software components and datasets. WINGS can reason about the constraints of the components and the characteristics of the data and propagate them through the workflow structure.

These approaches assist users by matching workflow components together. Instead, in our context, the workflows (or pipelines) are already available and need to be matched to relevant datasets. Conversely, relevant analysis workflows need to be recommended for a given dataset.

\subsection{Algorithm selection}

Other recommender systems aim at selecting specific algorithms for a given problem. For instance, the Oracle machine-learning toolkit~\cite{oracle} selects machine-learning algorithms and models for classiﬁcation and regression problems. Algorithm selection uses advanced machine learning techniques to automatically rank the best algorithms for a dataset. Model selection identifies the best hyperparameters to maximize a given prediction performance score. 

Another recent example is PennAI~\cite{la2021evaluating}, a platform that recommends suitable models given a supervised classification problem. The platform was evaluated on 165 classification problems. Results showed that matrix factorization-based recommendation systems outperform meta-learning methods.

In addition, Dyad ranking~\cite{tornede2019algorithm,tornede2020extreme} represents the algorithm and problem instance by a feature vector, and selects the best feature vectors using machine learning. The training dataset is a set of dyads ranked according to a specific preference relation. The dyad ranker learns using the neural-network-based PLNet~\cite{schafer2018dyad} algorithm. Results show that this approach outperforms many algorithm selectors while using less computation. 

All these approaches assume the existence of an objective function such as F1 score, accuracy or mean average error to compare algorithms or models that are known to apply to the problem. Instead, our goal is not to compare pipelines with each other but to predict if a given pipeline will work on a given dataset. From a methodological point of view, algorithm selection techniques are mostly content-based  while we will adopt a collaborative filtering approach.

\subsection{Finding tools and datasets in neuroimaging}

Several platforms have been developed to facilitate the sharing of tools and datasets 
in neuroimaging. For instance, NITRC~\cite{kennedy2016nitrc} provides a richly-annotated catalog of tools and datasets that can be processed in the NITRC computing environment. In OpenNeuro~\cite{markiewicz2021openneuro}, datasets complying to the BIDS data structure can be publicly shared and processed using BIDS apps.  Finally, the Neuroscience Information Framework (NIF~\cite{gardner2008neuroscience}) is a powerful search engine for neuroscience software and data. While all these platforms provide substantial services for data and tool sharing, they do not seem to include any system to assist users in the matching of pipelines and datasets. Matching relevant pipelines and datasets would also enable the automated triggering of data processing when new pipelines or datasets become available. 

\section{Materials and Methods}

\begin{figure*}[ht]
  \centering
  \includegraphics[width=\textwidth]{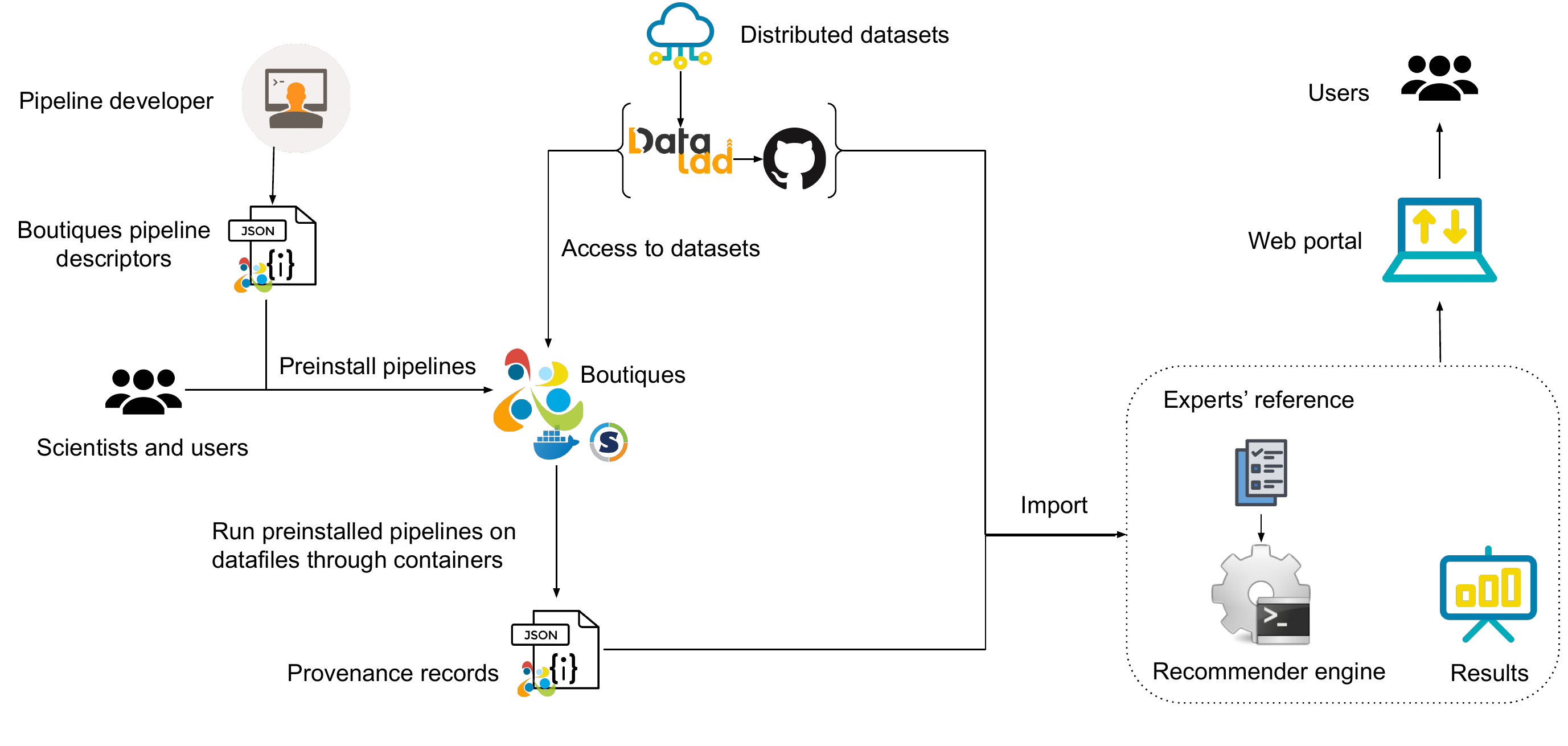}
  \caption{Overview of our recommender system}
  \label{fig:method}
  \end{figure*}

The goal of our recommender system is to predict if a data processing
pipeline will successfully run on a given dataset. Predictions are obtained
from provenance records created from previous pipeline executions.
Figure~\ref{fig:method} presents an overview of our system that is detailed
in the remainder of this section with our experimental methodology.


\subsection{Data processing pipelines} 
Consistently with our motivating use case, we focused on the 64
neuroscience pipelines available in the CONP as of October
2020. 
The available pipelines have 23 different tags which shows an overview of the type of data
and analyses supported by these pipelines. The most popular tags are ``neuroinformatics", ``mri", ``fmri", ``bioinformatics" and ``dmri". These pipelines are command-line
tools described by Boutiques descriptors~\cite{glatard2018boutiques} and
containerized using Docker or Singularity, which makes them portable across
a wide range of infrastructures, including local workstations, clusters,
and clouds~\cite{kiar2019serverless}. Each pipeline descriptor is stored on
the Zenodo research archive~\cite{https://doi.org/10.25495/7gxk-rd71} and identified by a permanent Digital Object
Identifier (DOI). Through the Boutiques command line, the pipelines can be
validated, installed, published, and executed. When the execution
completes, Boutiques creates a JSON provenance record containing a summary
of the execution process including the pipeline DOI, input and output file
hashes, parameter values, and exit code. Our recommendation system uses
such provenance records to track the outcome of pipeline executions on a
given dataset.

\subsection{Datasets} 

We used the 42 datasets available in the CONP as of October 2020,
representing a total of 2,807,267 files, 3,078~GB, and 6,330 subjects.
These datasets are coming from various sources distributed in Canada and abroad. The most frequent dataset keywords are ``brain imaging", ``human pain", ``MRI", ``neuroimaging" and ``structural MRI".
 Datasets are made available through DataLad~\cite{datalad2021},
a Git-based framework that provides a uniform and version-controlled view of
distributed storage.

A DataLad dataset is a particular type of Git repository that stores
 data files using git-annex~\cite{gitannex}. Git objects contain
hashes and URLs of the data files but not the data itself. With the DataLad
client, users can download specific files and track their versions. Using
DataLad, our system matches provenance records to particular datasets
through file hashes. In addition, specific files from a given dataset can
be downloaded on demand without having to download the entire dataset. 

 In some cases, minor adjustments such as file renamings or exclusions (on 8/22 of tested datasets) to the organization of BIDS datasets were performed to make them fully compliant with the standard specification and reduce the processing time of our experiments by excluding some subjects. We did not apply any data type conversion since in neuroimaging they are known to create issues when not done properly~\cite{li2016first}. 


\subsection{Expert reference}

To build an expert reference, we recruited 13 experts among graduate
students, software developers and data engineers at the Canadian Open
Neuroscience Platform. Since the number of pipeline-dataset pairs to
evaluate was beyond the amount that could reasonably be evaluated by a
human expert (2,688), we split our survey in two steps. In the first survey
(confidence survey), experts were asked to rate their knowledge and
confidence about each pipeline and dataset on a 4-level scale: no
knowledge, some knowledge, good knowledge, and expert knowledge. In the second
survey (prediction survey), experts were ask to predict the execution
outcome (success or failure) for all pipeline-dataset pairs in which they
had indicated good or expert knowledge in the first survey for both the
pipeline and the dataset. Both surveys included links to dataset and pipeline description pages on the CONP portal, such that the experts were able to consult their detailed descriptions. Survey forms are available on \href{https://github.com/big-data-lab-team/paper-pipelines-datasets-recommender/blob/master/data}{GitHub} for more information. Surveys happened between November 2020 and March 2021.

\subsection{Provenance records} 

Similar to pipeline descriptors, Boutiques provenance records can be
published to Zenodo, which makes them publicly and permanently accessible.
We created provenance records for each pipeline-dataset pair for which at least one expert predicted successful execution. To generate a provenance record we executed a pipeline using an invocation file including all required parameters for that pipeline. The instructions for generating the invocations are available through Boutiques for each pipeline. We mostly used default parameters and created minimal invocations, however, to execute some pipelines we requested domain experts' assistance to generate a working invocation file.
The provenance records were entered in a database together with the file hashes of all datasets retrieved using DataLad. From this database, we generated (pipeline, dataset, execution outcome) triplets to use in our recommender system.



\subsection{Recommender system}

Collaborative filtering predicts the rating of a given item (dataset in our
case) by a given user (pipeline in our case) from the ``utility
matrix" containing previous ratings~\cite{rajaraman2011mining}.
Two  approaches are commonly used: neighbor-based
methods~\cite{koren2015advances} and latent factor
models~\cite{koren2009matrix,bokde2015matrix}. Neighbor-based collaborative
filtering, also known as k-nearest neighbors, identifies
like-minded users or similar items based on the similarity of entries in
the utility matrix. In contrast, latent factor models, the method that we used, represent items and
users in a latent space obtained from a factorization of the
utility matrix $r$ in a user matrix $p$ and an item matrix $q$. The
factorization is learned by minimizing the least square error between the
available ratings and the ratings predicted by the factorization:
\begin{equation} \tag{1}
   \min_{q_{*},p_{*}} \sum_{(u,i) \epsilon \kappa }  (r_{ui}-q_{i}^{T}p_{u})^{^2}+\lambda \left ( \left \| q_{i}\right \|^2+\left \| p_{u}\right \|^2 \right )            \label{eq}
\end{equation}
where $\kappa$ is the set of user-item pairs for which $r_{ui}$, the rating of item
$i$ by user $u$, is available. Unknown elements of the utility matrix are
predicted by the dot product of the corresponding vectors in $q$ and $p$.
In our case, pipelines represent users, datasets represent items, execution
outcomes represent ratings, and $\kappa$ is the set of pipeline-dataset
pairs for which provenance records are available. Two minimization methods are
commonly used: stochastic gradient descent and alternating least squares
(ALS). We used ALS as implemented in the
Apache Spark Machine Learning library (spark.ml) version 3.1.2.

\section{Results}

We evaluated our approach in two different ways. First, 
we compared the expert predictions to 
real execution outcomes extracted from provenance records.
Second, we evaluated 
the accuracy of our recommender system through 10-fold cross validation. The data and code required to reproduce our results are available at  \url{https://github.com/big-data-lab-team/paper-pipelines-datasets-recommender}. 

\subsection{Expert predictions vs real executions}

\begin{figure*}
\centering
\begin{subfigure}{.49\textwidth}  \includegraphics[width=\columnwidth]{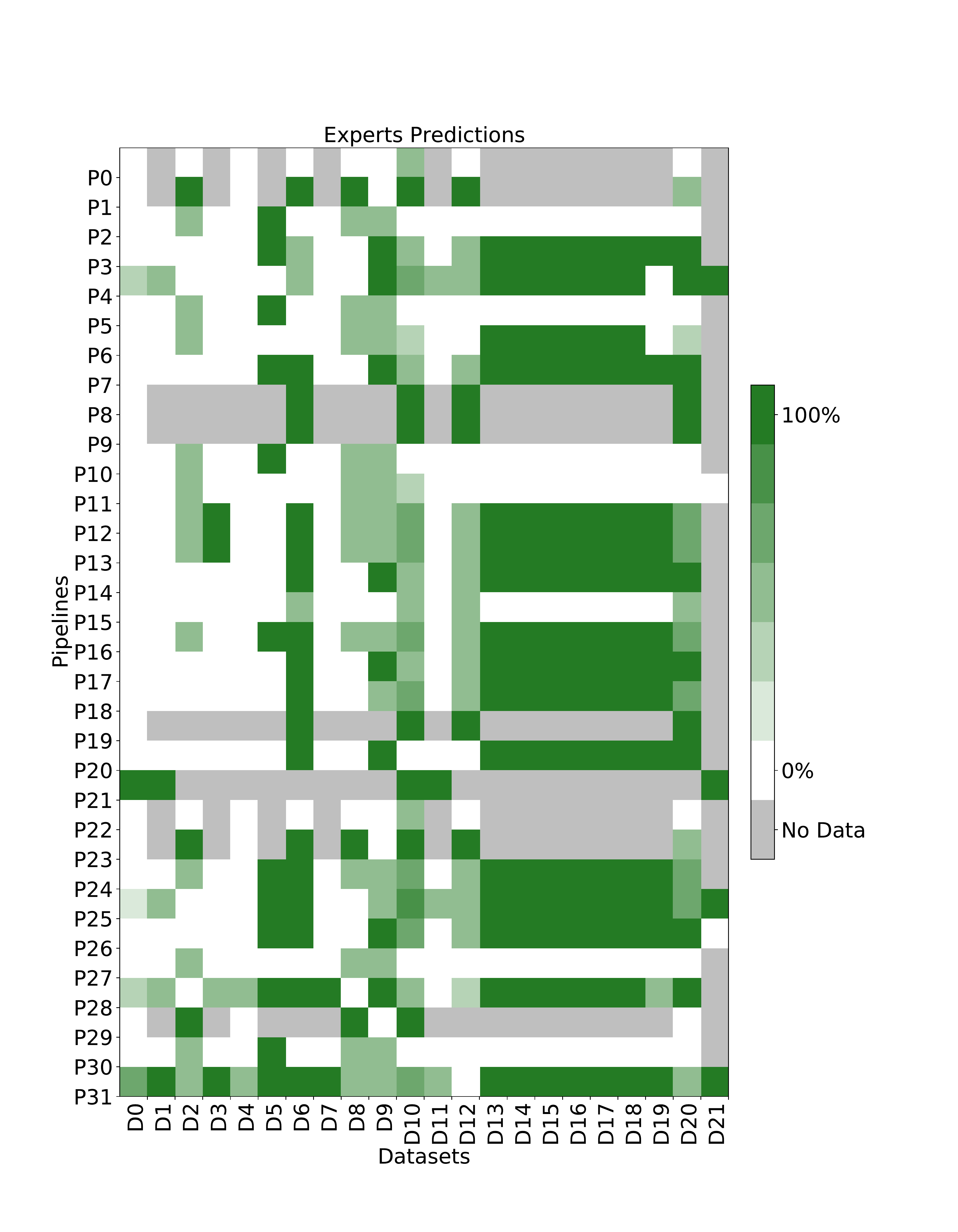}
  \caption{Experts predictions}
  \label{fig:experts_matrix}
\end{subfigure} \hfill
\begin{subfigure}{.49\textwidth}
  \includegraphics[width=\columnwidth]{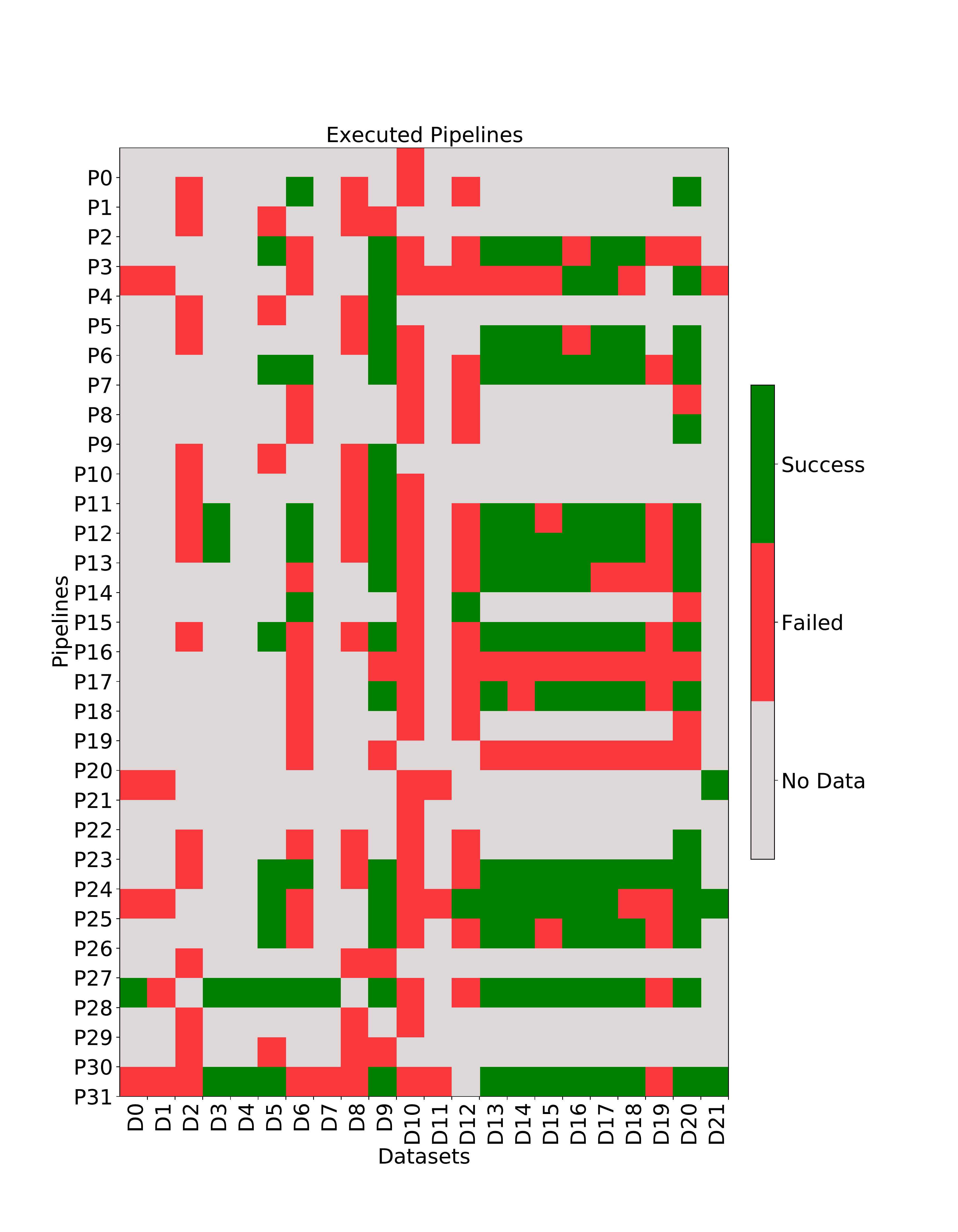}
  \caption{Pipeline outcomes}
  \label{fig:execution_matrix}
\end{subfigure}
\caption{Expert predictions vs real execution outcomes.}
\end{figure*}

Figure~\ref{fig:experts_matrix} shows expert predictions of
pipeline-dataset execution outcomes. The average number of expert predictions by pipeline-dataset pair was 1.39. Only the 32/64 pipelines for which at
least one dataset was predicted to be successfully processed by at least
one expert are represented.  Similarly, only the 22/42 datasets for which
at least one pipeline was predicted to be successfully executed by at least
one expert are represented. Pipeline and dataset names are reported in Tables~\ref{tab:pipeline_names} and~\ref{tab:data_names}. Entries in these tables are clickable for more information. Out of a total of 704 pipeline-dataset pairs,
the execution outcome of 37\% was predicted as failed by all the experts
(white cells), the outcome of 25\% was predicted as successful by all the
experts (dark green cells), 21\% were not known with enough confidence by
any expert (gray cells), and the outcome of the remaining ones was
predicted as successful by some experts and as failed by other experts. The
large fraction of pipeline-dataset pairs unknown to any expert reinforces
the motivation for an automated recommender system.

\begin{table}
    \centering
    \begin{tabular}{cc}
        \hline
        Index & Pipeline Name  \\
        \hline 
        $P_{0}$ & \href{https://doi.org/10.5281/zenodo.2621487}{ApplyTOPUP} \\
        $P_{1}$ & \href{https://doi.org/10.5281/zenodo.2650440}{ApplyWarp}\\
        $P_{2}$ & \href{https://doi.org/10.5281/zenodo.2566455}{BIDS App -- FSL Diffusion Preprocessing}\\
        $P_{3}$ & \href{https://doi.org/10.5281/zenodo.1484547}{BIDS App - FreeSurfer 6.0}\\
        $P_{4}$ & \href{https://doi.org/10.5281/zenodo.1895219}{BIDS App - fmriprep}\\
        $P_{5}$ & \href{https://doi.org/10.5281/zenodo.4457865}{BIDS App - ndmg}\\
        $P_{6}$ & \href{https://doi.org/10.5281/zenodo.1451001}{BIDS app example}\\
        $P_{7}$ & \href{https://doi.org/10.5281/zenodo.2636973}{ANTS Brain Extraction}\\
        $P_{8}$ & \href{https://doi.org/10.5281/zenodo.2640953}{ANTS Concat Transfo}\\
        $P_{9}$ & \href{https://doi.org/10.5281/zenodo.2634608}{ANTS Cortical Thickness}\\
        $P_{10}$ & \href{https://doi.org/10.5281/zenodo.2601876}{DTIFit}\\
        $P_{11}$ & \href{https://doi.org/10.5281/zenodo.3699595}{Dipy Tracking and Connectome Generation}\\
        $P_{12}$ & \href{https://doi.org/10.5281/zenodo.2597643}{FLIRT}	\\
        $P_{13}$ & \href{https://doi.org/10.5281/zenodo.2639849}{FNIRT}	\\
        $P_{14}$ & \href{https://doi.org/10.5281/zenodo.4043546}{FreeSurfer-Recon-all}\\
        $P_{15}$ & \href{https://doi.org/10.5281/zenodo.1450997}{FreeSurferPipelineBatch-CentOS7}\\
        $P_{16}$ & \href{https://doi.org/10.5281/zenodo.3267250}{FslBet601}	\\
        $P_{17}$ & \href{https://doi.org/10.5281/zenodo.1445789}{ICA\_AROMA}	\\
        $P_{18}$ & \href{https://doi.org/10.5281/zenodo.2602109}{MCFLIRT}	\\
        $P_{19}$ & \href{https://doi.org/10.5281/zenodo.3242714}{MRIQC}	\\
        $P_{20}$ & \href{https://doi.org/10.5281/zenodo.1450995}{PreFreeSurferPipelineBatch}	\\
        $P_{21}$ & \href{https://doi.org/10.5281/zenodo.4010734}{SPARK (stage 1 of 3)}	\\
        $P_{22}$ & \href{https://doi.org/10.5281/zenodo.2621482}{TOPUP}	\\
        $P_{23}$ & \href{https://doi.org/10.5281/zenodo.3899496}{fsl\_anat}	\\
        $P_{24}$ & \href{https://doi.org/10.5281/zenodo.1482743}{fsl\_bet}	\\
        $P_{25}$ & \href{https://doi.org/10.5281/zenodo.1494308}{fsl\_fast}	\\
        $P_{26}$ & \href{https://doi.org/10.5281/zenodo.1494312}{fsl\_first}	\\
        $P_{27}$ & \href{target: https://doi.org/10.5281/zenodo.1450991}{fsl\_probtrackx2}	\\
        $P_{28}$ & \href{https://doi.org/10.5281/zenodo.4472771}{fslstats}	\\
        $P_{29}$ & \href{https://doi.org/10.5281/zenodo.2566443}{mask2boundary}\\
        $P_{30}$ & \href{https://doi.org/10.5281/zenodo.1450999}{ndmg}	\\
        $P_{31}$ & \href{https://doi.org/10.5281/zenodo.3308620}{oneVoxel}\\
        \hline\\
        &
    \end{tabular}
    \caption{Tested pipelines}
    \label{tab:pipeline_names}
\end{table}



\begin{table}
    \centering
    \begin{tabular}{cc}
        \hline
        Index & Dataset Name  \\
        \hline 
        $D_{0}$ & \href{https://portal.conp.ca/dataset?id=projects/bigbrain-datalad}{BigBrain} \\[5pt]
        $D_{1}$ & \href{https://portal.conp.ca/dataset?id=projects/BigBrain_3DClassifiedVolumes}{BigBrain\_3DClassifiedVolumes}\\[5pt]
        $D_{2}$ & \parbox{3cm}{\href{https://portal.conp.ca/dataset?id=projects/Comparing_Perturbation_Modes_for_Evaluating_Instabilities_in_Neuroimaging__Processed_NKI_RS_Subset__08_2019_}{Comparing\_Perturbation\_Modes       \_for\_Evaluating\_Instabilities
        \_in\_Neuroimaging\_\_Processed\_NKI
        \_RS\_Subset\_\_08\_2019\_}}\\[5pt]
                &\\
        $D_{3}$ & \href{https://portal.conp.ca/dataset?id=projects/Khanlab/BigBrainHippoUnfold}{BigBrainHippoUnfold}\\ [5pt]      
        $D_{4}$ & \href{https://portal.conp.ca/dataset?id=projects/Khanlab/BigBrainMRICoreg}{BigBrainMRICoreg}\\[5pt]
        $D_{5}$ & \href{https://portal.conp.ca/dataset?id=projects/Khanlab/HCPUR100-Template}{HCPUR100-Template}\\[5pt]
        $D_{6}$ & \parbox{3cm}{\href{https://portal.conp.ca/dataset?id=projects/Learning_Naturalistic_Structure__Processed_fMRI_dataset}{Learning\_Naturalistic\_Structure\\
        \_Processed\_fMRI\_dataset}}\\[5pt]
                 &\\
        $D_{7}$ & \parbox{3cm}{\href{https://portal.conp.ca/dataset?id=projects/MRI_and_unbiased_averages_of_wild_muskrats__Ondatra_zibethicus__and_red_squirrels__Tamiasciurus_hudsonicus_}{MRI\_and\_unbiased\_averages\\
        \_of\_wild\_muskrats\_\\
        \_Ondatra\_zibethicus\_\\
        \_and\_red\_squirrels\_\\
        \_Tamiasciurus\_hudsonicus\_}} \\[5pt]
                 &\\        
                 
        $D_{8}$ & \parbox{3cm}{\href{https://portal.conp.ca/dataset?id=projects/Numerically_Perturbed_Structural_Connectomes_from_100_individuals_in_the_NKI_Rockland_Dataset}{Numerically\_Perturbed\_Structural\\
                \_Connectomes\_from\_100\_individuals\_ \\in\_the\_NKI\_Rockland\_Dataset}}\\[5pt]
                &\\
        $D_{9}$ & \href{https://portal.conp.ca/dataset?id=projects/SIMON-dataset}{SIMON-dataset}\\
        $D_{10}$ & \href{https://portal.conp.ca/dataset?id=projects/cneuromod}{cneuromod}\\
        $D_{11}$ & mm\_neo\_atlas\\
        $D_{12}$ & \href{https://portal.conp.ca/dataset?id=projects/multicenter-phantom}{multicenter-phantom}\\
        $D_{13}$ & \href{https://portal.conp.ca/dataset?id=openpain-BrainNetworkChange_Mano}{openpain/BrainNetworkChange\_Mano}\\
        $D_{14}$ & \href{https://portal.conp.ca/dataset?id=openpain-cbp_resting}{openpain/cbp\_resting}\\
        $D_{15}$ & \href{https://portal.conp.ca/dataset?id=openpain-placebo_1}{openpain/placebo\_1}\\
        $D_{16}$ & \href{https://portal.conp.ca/dataset?id=openpain-placebo_predict_tetreault}{penpain/placebo\_predict\_tetreault}\\
        $D_{17}$ & \href{https://portal.conp.ca/dataset?id=projects/openpain/subacute_longitudinal_study}{openpain/subacute\_longitudinal\_study}\\
        $D_{18}$ & \href{https://portal.conp.ca/dataset?id=openpain-thermal}{openpain/thermal}\\
        $D_{19}$ & \href{https://portal.conp.ca/dataset?id=projects/preventad-open}{preventad-open}\\
        $D_{20}$ & \href{https://portal.conp.ca/dataset?id=projects/preventad-open-bids}{preventad-open-bids}\\
        $D_{21}$ & \href{https://portal.conp.ca/dataset?id=projects/visual-working-memory}{visual-working-memory}\\

        \hline \\
        &
    \end{tabular}
    \caption{Tested datasets}
    \label{tab:data_names}
\end{table}

\begin{figure}
  \includegraphics[width=\columnwidth]{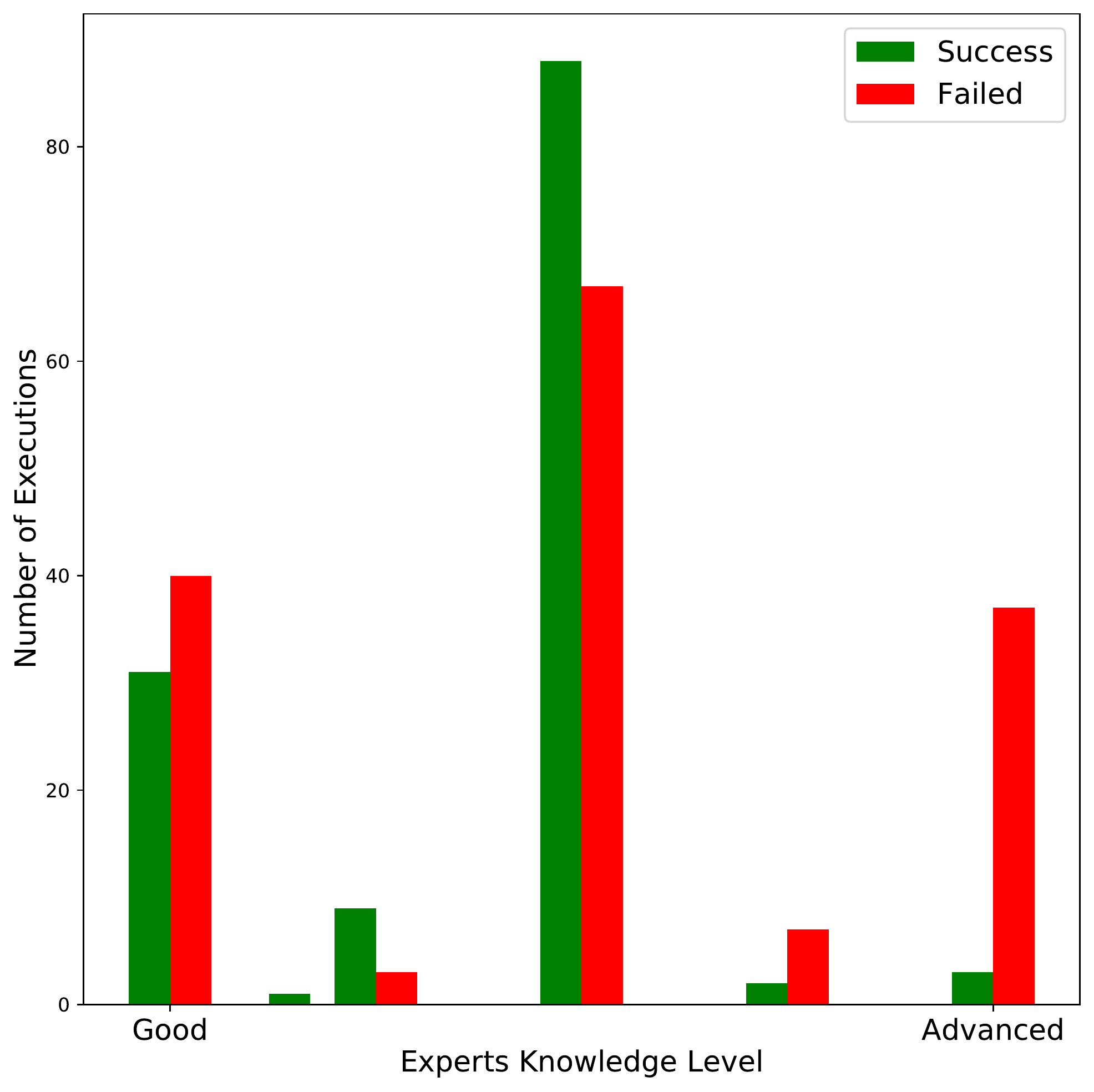}
  \caption{Expert confidence by actual execution outcome.}
  \label{fig:confidence_swarm}
\end{figure}

Figure~\ref{fig:execution_matrix} shows the actual execution outcome for
all the pipeline-dataset pairs for which at least one expert predicted a
successful execution outcome (green cells in
Figure~\ref{fig:experts_matrix}). Out of 288 executed pairs, 134   were
successful and 154 failed. 
Important discrepancies are observed between expert predictions and actual
executions. Overall, 53\% of the executions that were predicted successful by
at least one expert failed in reality (red cells in
Figure~\ref{fig:execution_matrix}). In addition, the average expert
confidence was found to be significantly higher for failed executions than
for successful ones (p \textless 0.002, Figure
~\ref{fig:confidence_swarm}), which is unexpected. Therefore, expert predictions
seem to be largely unreliable. Note that we used the experts' predictions as a baseline for prediction performance comparison rather than a ground truth on the execution outcome of a given pipeline on a given dataset. 

Many practical reasons explain the observed discrepancy between expert
predictions and pipeline executions (Table~\ref{tab1}). First, some
datasets did not match the format required by the tested pipeline. For
instance, $P_{26}$ (fsl-first) requires anatomical images in the NIfTI file
format, however, some datasets such as $D_{12}$ (multicenter-phantom)
contain anatomical images in the MINC format.

\begin{table}[htbp]
 
    \begin{center}
        \begin{tabular}{cc}
            \hline
            \textbf{Failure reason} & \textbf{Fraction of failed executions}  \\
            \hline
            \textbf{File format not supported}    & 31\% \\
            \textbf{Type D pre-processing required}    & 18\% \\
            \textbf{Dataset not available}         & 38.5\% \\
            \textbf{Other}       & 12.5\% \\
            \hline
        \end{tabular}
           \caption{Execution failure causes}
        \label{tab1}
    \end{center}
\end{table}

In addition, five pipelines ($P_{8}$, $P_{17}$, $P_{19}$, $P_{20}$ and
$P_{27}$) failed due to unresolved pre-processing requirements. We
identified four types of dependencies between pipelines. Type A refers to pipelines such
as $P_{24}$ (fsl-bet) or $P_{23}$ (fsl-anat) that can be executed directly
on the tested dataset. Type B refers to pipelines that require the tested
dataset as well as the results of the application of another pipeline on
the tested dataset. For example, $P_{31}$ (oneVoxel) requires a binary mask
for its input image that is created by another pipeline. Type C refers
to pipelines requiring inputs from more than one dataset. For instance, $P_7$
(ANTS Brain Extraction) and $P_9$ (ANTS Cortical Thickness) require
external templates and segmentations obtained outside of the dataset. Type
D refers to pipelines that process data derived from the tested dataset but not
the tested dataset directly.  For instance, $P_{27}$ (fsl-probtrackx2) performs
probabilistic tractography on the output of bedpostx, a pipeline that
no expert predicted to run successfully on any dataset. In a Type D
configuration, the pipeline is considered to not successfully execute on
the dataset. 

In addition, 4 datasets were not available for download in CONP, due to
various issues. For example, $D_{10}$ (CNeuromod) is currently not
downloadable in CONP due to technical issues.
Finally, a set of executions failed for other reasons including issues in
Boutiques pipeline descriptors or corrupted datasets.

Overall, experts seem to have neglected such practical failure reasons. In
general, experts tend to rely on their semantic understanding of the
interactions between pipelines and datasets (for instance, a given pipeline
may operate on fMRI data), while in practice, pipeline executions depend on
the lower-level syntactical and infrastructural details mentioned previously.



\subsection{Recommender system evaluation} 

\begin{figure}
\centering
  \includegraphics[width=\columnwidth]{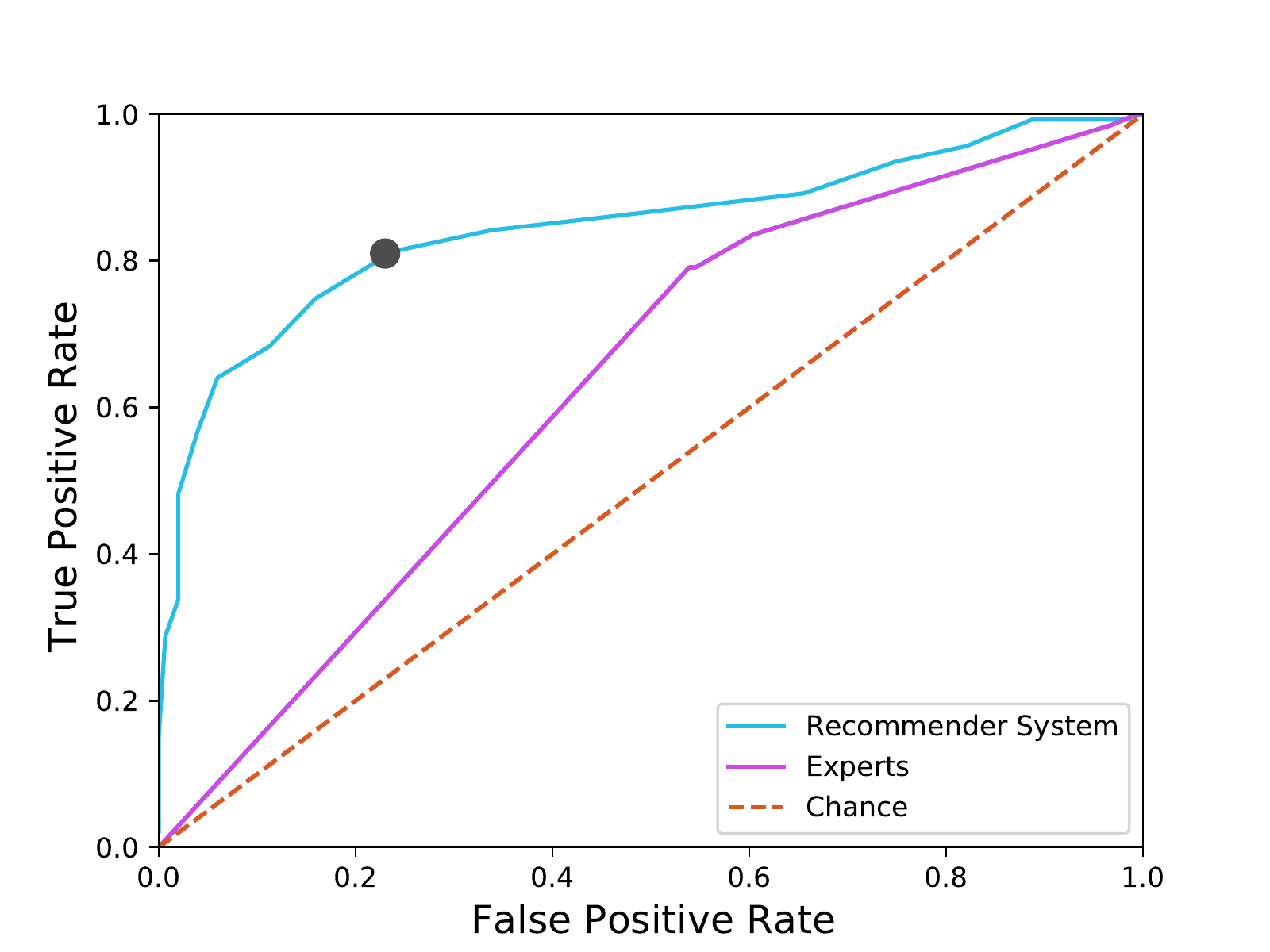}
  \caption{ROC curves of experts and recommender system predictions.}
  \label{fig:roc-curve}
\end{figure}

We evaluated the latent-factor model using 10-fold cross validation on the
pipeline execution matrix in Figure~\ref{fig:execution_matrix}. We varied
the threshold used to round predicted values to 1 (failed execution) or 2
(successful execution), resulting in the Receiver Operating Characteristic
(ROC) curve  in Figure~\ref{fig:roc-curve}. We obtained the ROC curve of
experts predictions by predicting execution outcomes using various
thresholds in the fraction of experts predicting successful execution. 

The area under curve (AUC) of our recommender system was 0.83, showing that
our model is significantly better than chance (AUC=0.5) and expert
predictions (AUC=0.63). For instance, given a rounding threshold of 1.2
(black dot in the ROC curve), out of 10 pipelines
recommended by our system for a particular dataset, 8 would be applicable to
the dataset while only 2 would not. This good performance was expected to some degree given that the pipeline execution matrix in
Figure~\ref{fig:execution_matrix} bears some sort of structure. 
A more random utility matrix would obviously be more difficult to predict.

\section{Discussion}


The performance of the proposed recommender system is substantially higher
than chance and expert predictions. Predicting the successful execution
outcome of a pipeline on a given dataset is a difficult task for a human expert as it requires a comprehensive knowledge and understanding of the technical infrastructure, pipeline syntactical requirements, analysis types, data formats, and data semantic types. In practice, it is common for human experts to only master some parts of this environment. For example,
the successful processing of BIDS datasets by BIDS applications requires
datasets to pass BIDS validation, which can hardly be guessed from a
high-level overview of the dataset. 
In addition, pipelines or data
transfers may fail for technical reasons unknown to the experts. Therefore, automated
recommender systems based on provenance records have a strong added value
compared to human recommendations. 

Our experiments were conducted using one of the largest data and pipeline
sharing platform in neuroscience. Other platforms such as
NITRC~\cite{kennedy2016nitrc} contain larger collections of
pipelines and datasets, but they are not available through a consistent
interface such as Boutiques and DataLad, which would make
such an experiment hardly feasible there.

The described system architecture could potentially scale widely beyond the specific
context of CONP, as it only relies on file hashes and therefore does not
require data sharing agreements or extensive data storage. In addition, the
recommender system could leverage provenance records produced by multiple
platforms provided that they are shared in some way.
DataLad would detect possible duplication among datasets through file hashes. 
The framework is also expected to apply to other disciplines than neuroimaging
although changes in the technical context --- such as datasets being stored in 
databases instead of files --- may require adaptations.

In production conditions, the recommender system would rely on provenance
records of pipeline executions launched by arbitrary users. While this would increase
the amount of data available, potentially resulting in more accurate
recommendations, it would also come with challenges. For instance, while
our framework models the execution outcome of a given pipeline-dataset pair
as a binary variable (success or failure), different execution outcomes may
be produced for a given pipeline-dataset pair, due to different
parametrizations or analysis types. Besides, analyses launched by less
experimented users may produce misleading provenance records. We anticipate
that the recommender could be configured to use implicit feedback to address 
this issue~\cite{hu2008collaborative}.

The successful execution of a pipeline on a given dataset 
does not necessarily imply that results are meaningful. Relying exclusively 
on execution exit statuses therefore requires that users producing execution records mostly execute meaningful experiments, which may not always be the case. Taking into account the popularity of
the datasets derived from a given provenance record in the recommendations might help address this issue.



\section{Conclusion}

Collaborative filtering predicts the execution outcome of a given pipeline
on a given dataset with usable accuracy (AUC=0.83) in the context of the
Canadian Open Neuroscience Platform. The performance achieved by our system
outperforms
human expert recommendations, presumably due to syntactical and
infrastructural factors neglected by human experts. 
Future work will focus on the deployment of such a system in production conditions, which will require dealing with less reliable provenance records.


The framework could be extended by considering pipelines and datasets 
at a finer granularity. Pipelines can often be used in different ways 
depending on their parametrization. Different parametrizations could 
be identified in the provenance records and recommended accordingly 
for specific datasets. Besides, datasets often consist of multiple 
sub-parts corresponding to different subjects or data types. A recommender 
system could be designed to recommend analyses for such sub-parts, 
resulting in more specific recommendations.






\bibliographystyle{plain}
\bibliography{bibliography}

\end{document}